\documentclass[a4paper,11pt]{article}
\pdfoutput=1 

\usepackage{jheppub} 

\usepackage[T1]{fontenc} 
\usepackage{blkarray}
\usepackage{relsize}
\usepackage{slashed}
\usepackage{filecontents}
\usepackage{amsthm}
\usepackage{mathtools}

\def \barlambda {\bar{\lambda}}
\def \barmu {\bar{\mu}}

\makeatletter
\DeclareFontFamily{OMX}{MnSymbolE}{}
\DeclareSymbolFont{MnLargeSymbols}{OMX}{MnSymbolE}{m}{n}
\SetSymbolFont{MnLargeSymbols}{bold}{OMX}{MnSymbolE}{b}{n}
\DeclareFontShape{OMX}{MnSymbolE}{m}{n}{
    <-6>  MnSymbolE5
   <6-7>  MnSymbolE6
   <7-8>  MnSymbolE7
   <8-9>  MnSymbolE8
   <9-10> MnSymbolE9
  <10-12> MnSymbolE10
  <12->   MnSymbolE12
}{}
\DeclareFontShape{OMX}{MnSymbolE}{b}{n}{
    <-6>  MnSymbolE-Bold5
   <6-7>  MnSymbolE-Bold6
   <7-8>  MnSymbolE-Bold7
   <8-9>  MnSymbolE-Bold8
   <9-10> MnSymbolE-Bold9
  <10-12> MnSymbolE-Bold10
  <12->   MnSymbolE-Bold12
}{}

\let\llangle\@undefined
\let\rrangle\@undefined
\DeclareMathDelimiter{\llangle}{\mathopen}%
                     {MnLargeSymbols}{'164}{MnLargeSymbols}{'164}
\DeclareMathDelimiter{\rrangle}{\mathclose}%
                     {MnLargeSymbols}{'171}{MnLargeSymbols}{'171}
\makeatother

\allowdisplaybreaks

\title{\boldmath A Pure Spinor Twistor Description of \\ Ambitwistor Strings}

\author[a,b]{Diego Garc\'ia Sep\'ulveda}
\author[c]{and Max Guillen}

\affiliation[a]{Perimeter Institute for Theoretical Physics, Waterloo, ON N2L 2Y5, Canada}
\affiliation[b]{
ICTP South American Institute for Fundamental Research \\
Instituto de F\'isica Te\'orica, UNESP-Universidade Estadual Paulista \\
R. Dr. Bento T. Ferraz 271, Bl. II, S\~ao Paulo 01140-070, SP, Brazil}
\affiliation[c]{Department of Physics and Astronomy,\\
Uppsala University, 75108 Uppsala, Sweden}

\emailAdd{diego.garcia@unesp.br}
\emailAdd{max.guillen@physics.uu.se}

\abstract{We present a novel ten-dimensional description of ambitwistor strings. This formulation is based on a set of supertwistor variables involving pure spinors and a set of constraints previously introduced in the context of the $D=10$ superparticle following a ten-dimensional twistor-like construction introduced by Berkovits. We perform a detailed quantum-mechanical analysis of the constraint algebra, we show that the corresponding central charges vanish, and after considering a convenient gauge fixing procedure, physical states are found. Vertex operators are explicitly constructed and, by noticing a relation with the standard pure spinor formalism, scattering amplitudes are shown to correctly describe $D=10$ super-Yang-Mills interactions. As in other ambitwistor string models, amplitudes are found to be localized on the support of the scattering equations, and thus this work provides a bridge between Berkovits' construction and the Cachazo-He-Yuan formulae. After extending the pure spinor twistor transform to include an additional supersymmetry, our results are immediately generalized to Type IIB supergravity.

}

\begin{document} 
\maketitle

\flushbottom

\section{Introduction} \label{introduction}

After the remarkable discovery of Cachazo, He, and Yuan (CHY) \cite{Cachazo:2013hca, Cachazo:2013iea, Cachazo:2014xea} of general compact formulae for tree-level amplitudes as integrals over the space of punctured Riemann spheres localized over the so-called \textit{scattering equations}, it became an immediate issue how to consider fermions or supersymmetry into the formalism. Shortly after these original findings, ambitwistor strings \cite{Mason:2013sva, Berkovits:2013xba} were found to naturally give rise to the CHY formulae and provided a natural framework to consider supersymmetric generalizations of the latter. In this regard, compact amplitudes formulae with supersymmetry have been constructed in four, six, and ten-eleven dimensions in \cite{Geyer:2014fka, Geyer:2018xgb, Geyer:2019ayz} making use of twistor variables instead of the standard superspace variables.

On another line of developments, a novel formulation of the ten-dimensional massless superparticle in terms of twistor-like variables was introduced by the authors in a complementary work \cite{Sepulveda:2020kjc}. This formulation was found by looking for a first-principles description of a twistor-like construction in ten dimensions introduced by Berkovits in \cite{Berkovits:2009by}, in which a set of ``pure spinor twistor'' variables 
\begin{equation}\label{ztwistors}
    \mathcal{Z}^{I} = (\lambda^{\alpha}, \mu_{\alpha}, \Gamma^{m}), \ \ \bar{\mathcal{Z}}_{I} = (\barmu_{\alpha}, - \barlambda
^{\alpha}, \bar{\Gamma}^{m}), 
\end{equation}
where $m=1,\ldots,10$, $\alpha=1,\ldots,16$, fulfilling
\begin{equation} \label{definingconstraints}
    (\lambda \gamma^{m} \lambda) = 0, \ \ \ \lambda \mu = 0, \ \ \ (\lambda \gamma^{mn} \mu) + 4\Gamma^{m} \Gamma^{n} = 0, \ \ \ (\lambda \gamma_{m})_{\alpha} \Gamma^{m}=0
\end{equation}
were used in an attempt to generalize standard four-dimensional twistor constructions \cite{Witten:2003nn, Berkovits:2004hg, Roiban:2004vt, Britto:2004ap, Britto:2005fq} to ten dimensions, with pure spinors taking the role of higher dimensional twistors, which is a natural proposal as argued in \cite{Hughston1, Hughston2, Hughston3, Berkovits:2004bw, Boels:2009bv}. \pagebreak

In this work we will present an ambitwistor worldsheet theory based on the previously mentioned description of the superparticle \cite{Sepulveda:2020kjc}. The worldsheet theory completes the physical realization of \cite{Berkovits:2009by} when arbitrary interactions are considered, and there are resemblances with the previously mentioned models \cite{Geyer:2014fka, Geyer:2018xgb, Geyer:2019ayz}. For instance, both constructions use a twistorial representation that makes supersymmetry manifest, and as we shall see, vertex operators take similar forms. The ambitwistor string considered here is constructed by replacing time derivatives by antiholomorphic derivatives $\bar{\partial}$ and the worldline by a Riemann sphere, in accordance with the ideas of \cite{Mason:2013sva} to construct ambitwistor worldsheet actions. As in the superparticle, the worldsheet variables by definition will be required to satisfy \eqref{definingconstraints}, and the system will be subjected to a set of constraints

\begin{align}
    B & \coloneqq (\lambda\gamma^{m}\bar{\lambda})\bar{\Gamma}_{m} - (\bar{\lambda}\gamma^{m}\bar{\lambda})\Gamma_{m}, \label{Bconstraint}\\[1.0ex]
    J & \coloneqq \barmu_{\alpha} \lambda^{\alpha} - \barlambda^{\alpha} \mu_{\alpha} + \bar{\Gamma}^{m} \Gamma_{m} + j_{c} \label{Jconstraintintro}, \\[1.0ex]
    \tilde{\mathcal{M}}^{abc} & \coloneqq (\lambda \gamma^{[a}) N^{bc]} + \frac{1}{12}(\tilde{q} \gamma^{abc} \tilde{q}) \label{splconstraint}
\end{align}
apart from the corresponding Virasoro constraint $T=0$. Here, $j_{c}$ is a quantum correction to be determined, and $a,b,c=1, \ldots, 5$ are $SU(5)$ indices. Notice that the constraints $B$ and $J$ have both been considered previously in \cite{Berkovits:2009by} and emerged naturally in the context of the superparticle, but the importance of the constraint $\tilde{\mathcal{M}}^{abc}$ -the independent 
components of the super-Pauli-Lubanski three-form- to properly describe the degrees of freedom of the $D=10$ Brink-Schwarz superparticle in the twistor framework was pointed out in \cite{Sepulveda:2020kjc}. The worldsheet action so constructed would then lead to an interesting resolution to a conjecture proposed by Berkovits in \cite{Berkovits:2009by}; the worldsheet action, rather than related at first sight to the standard superstring, would be related to an ambitwistor string theory.

We construct a heterotic version of the model just introduced by coupling the system to a current algebra. Remarkably, the whole set of constraints $(B,J,T,\tilde{\mathcal{M}}^{abc})$ gives rise to an anomaly-free worldsheet model when the current algebra central charge is 16, as in the $E_{8}\times E_{8}$ or $SO(32)$ heterotic superstrings. Analogously, the Type IIB version constructed out of a simple extension of \eqref{ztwistors} and \eqref{definingconstraints} will also present a vanishing central charge.

In \cite{Sepulveda:2020kjc}, the pure spinor twistor formulation of the $D=10$ superparticle was found through a field redefinition of a superparticle model developed by Berkovits in \cite{Berkovits:1990yc}. The corresponding ambitwistor string constructed from \cite{Berkovits:1990yc} has been studied by Berkovits, Mason, and one of the authors in \cite{Berkovits:2019bbx}, where it was shown that in light-cone gauge the model is equivalent to the light-cone RNS ambitwistor string. Due to the close relation between the latter model and the ambitwistor string constructed in this work from \cite{Sepulveda:2020kjc}, we begin warming-up by finding the BRST operator for \cite{Berkovits:2019bbx} before constructing the worldsheet model with pure spinor twistor variables. This provides a first step onto covariant quantization of the ambitwistor string in \cite{Berkovits:2019bbx}.

In virtue of \eqref{definingconstraints}, the operator product expansions (OPEs) satisfied by the pure spinor twistor worldsheet variables do not correspond to those of a free theory. We thus need to resort to the tools of interacting two-dimensional conformal field theories (2D CFTs) as outlined in \cite{DiFrancesco:1997nk, Bais:1987dc} to compute OPEs between different operators of interest. Notice that the same set of tools have been used in \cite{Oda:2005sd, Oda:2007ak} in the context of the standard pure spinor formalism in order to reproduce the corresponding OPEs as originally found in \cite{Berkovits:2000fe}. We set up an analogous construction for the model developed here, and we find the corresponding expressions for the stress-energy tensor $T$, the Lorentz generator $N^{mn}$, and the projective weight operator $J$, all of which develop corrections similarly as in \cite{Oda:2005sd, Oda:2007ak}. In particular, we find that $J$ develops anomalies at quantum level. 

As we will see, the model we present in this work has many similarities with the ordinary pure spinor ambitwistor string. Indeed, to make these similarities transparent we shall fix only the $B$-constraint  \eqref{Bconstraint} and leave other symmetries unfixed, from which a corresponding BRST operator can be constructed. Moreover, the latter procedure will also be instrumental to see how the momentum conservation delta function arises in the pure spinor twistor model.

As in standard superstring theory, we write scattering amplitudes as correlation functions of vertex operators in pictures $0$ and $-1$. The picture $-1$ vertex operators are obtained from the corresponding superparticle wavefunction \cite{Sepulveda:2020kjc}, first considered in the original work \cite{Berkovits:2009by}, and we construct picture $0$ vertex operators through a picture raising operation. Integrated vertex operators share similar properties as those of other ambitwistor models \cite{Mason:2013sva, Berkovits:2013xba, Geyer:2014fka, Geyer:2018xgb, Geyer:2019ayz, Berkovits:2019bbx}, being localized in a set of delta functions that lead to momentum conservation and that localize over the scattering equations $k_{i} \cdot P_{(z_{i})} = 0$ \cite{Cachazo:2013hca, Cachazo:2013iea, Cachazo:2014xea}. Vertex operators are written in terms of integrals over ``auxiliary'' $SU(5)$ vectors $\epsilon^{\dot{a}}, \bar{\epsilon}_{\dot{a}}$ that are introduced by noticing that the momentum $P^{m} = (\lambda \gamma^{m} \barlambda)$ satisfies a ``twistor-like constraint'' $(\lambda \gamma_{m})_{\alpha}P^{m} =0$ \cite{Berkovits:2015yra} and it is then left invariant under an $SU(5)$ subgroup of the complexified ten-dimensional Wick-rotated spacetime. We then associate an additional $SU(5)$ index to our variables which allows us to write down explicit expressions for the pictures $0$ and $-1$ vertex operators.

The amplitude prescription is a standard proposal consisting of integrating over all independent components of our fields and modding out by the killing vectors redundancies. As usual, we have to mod out by $\textrm{SL}(2,\mathbb{C})$ arising from reparametrizations, but we will also need to mod out by the $\textrm{GL}(1)$ associated to \eqref{Jconstraintintro} and the killing vectors associated to \eqref{splconstraint}. As we shall see, this prescription will turn out to be related to the corresponding prescription found in the standard pure spinor ambitwistor model \cite{Berkovits:2013xba}: the measure is tantamount to that of the pure spinor formalism $\langle (\lambda \gamma^{m}\theta) (\lambda \gamma
^{n}\theta) (\lambda \gamma^{p}\theta) (\theta \gamma_{mnp} \theta)\rangle = 1$, and both unintegrated as well as integrated vertex operators can be related, on the support of the incidence relations and the delta functions appearing in the vertex operators, to their counterparts in the standard pure spinor formalism. We conclude from this observation that our amplitude prescription indeed gives the correct correlators.

This work is organized as follows: In section \ref{section2} we review the ambitwistor model arising from Berkovits' superparticle model \cite{Berkovits:1990yc} and write the covariant BRST operator left as an open problem from \cite{Berkovits:2019bbx}. Sections \ref{section3} and \ref{section4} contain the main results of this paper. In section \ref{section3} we construct the pure spinor twistor ambitwistor model from the superparticle formulation of \cite{Sepulveda:2020kjc}, we construct the corresponding BRST operator, and write down vertex operators in pictures $-1$ and $0$. In section \ref{section4} we discuss the scattering amplitudes prescription along with its relation to the standard pure spinor formalism. We conclude in section \ref{section5} with discussions and some directions for further research. In Appendix \ref{AppendixA} we provide a quick review to the tools used to compute OPEs in interacting 2D CFTs.

\section{The Non-Pure Spinor Description of Ambitwistor Strings} \label{section2}

In this section we consider an ambitwistor model first described in \cite{Berkovits:2019bbx}, where it was shown that the model correctly reproduced the spectrum of the RNS ambitwistor string. However, the analysis was performed in light-cone gauge and a covariant analysis was left as an open problem owing to the reducibilities present in the constraints. Due to the relation between this model and the one we will describe in section \ref{section3} (via a field redefinition explained in \cite{Sepulveda:2020kjc}), we warm-up in this section constructing the BRST operator for the model described in \cite{Berkovits:2019bbx}, which in principle provides the means for a covariant analysis.

We start defining $\Lambda^{\alpha}$, $\Omega_{\beta}$ to be ten-dimensional Majorana-Weyl spinors of opposite chirality, and $\psi^{m}$ to be a ten-dimensional fermionic vector. We use letters from the beginning/middle of the Greek/Latin alphabet to denote ten-dimensional spinor/vector indices. The relation between these variables and the standard superspace variables is simply given by:
\begin{equation}
\Omega_{\alpha} = (\gamma_{m}\Lambda)_{\alpha}X^{m} + \psi^{m}(\gamma_{m}\theta)_{\alpha},  \ \ \ \  \psi^{m} = (\Lambda\gamma^{m}\theta),
\end{equation}
where $(\gamma^{m})_{\alpha\beta}$, $(\gamma^{m})^{\alpha\beta}$ are the ten-dimensional Pauli matrices satisfying the standard Dirac algebra $(\gamma^{(m})_{\alpha \gamma}(\gamma^{n)})^{\gamma \beta} = \eta^{mn}\delta_{\alpha}^{\beta}$ as well as the special identity $(\gamma^{m})_{(\alpha\beta}(\gamma_{m})_{\delta)\epsilon} = 0$ valid in ten dimensions.

The former variables were originally introduced in \cite{Berkovits:1990yc} in order to construct a model of the ten-dimensional massless superparticle with only first-class constraints. The corresponding heterotic ambitwistor string was later introduced in \cite{Berkovits:2019bbx}, with an action given by
\begin{equation}\label{heteroriginalaction}
    S = \int \Big[ \frac{1}{2} \Lambda \bar{\partial} \Omega - \frac{1}{2} \Omega \bar{\partial} \Lambda - \frac{1}{2} \psi_{m} \bar{\partial} \psi^{m} + \xi_{\alpha}G^{\alpha} + \kappa T_{F} \Big] + S_{J},
\end{equation}
where $S_{J}$ stands for a current algebra and $\xi_{\alpha}$ and $\kappa$ are Lagrange multipliers enforcing a set of constraints with two levels of reducibility:
\begin{align}
    & T_{F} \coloneqq (\Lambda \gamma^{m} \Lambda) \psi_{m}, \label{TFconstraint} \\[1.0ex]
    & G^{\alpha} \coloneqq (\Lambda \gamma^{m} \Lambda) (\gamma_{m} \Omega)^{\alpha} - 2 \Lambda^{\alpha} (\Lambda \Omega) + \psi^{m} \psi^{n} (\gamma_{m} \gamma_{n} \Lambda)^{\alpha}= 0, \label{Gconstraint} \\[1.0ex]
    & H^{m} \coloneqq (\Lambda\gamma^{m}G) - 2\psi^{m}T_{F} = 0, \label{one-reducibility} \\[1.0ex]
    \label{two-reducibility} 
    &(\Lambda \gamma^{m} \Lambda) H_{m} = 0.
\end{align}
The stress-energy tensor for this worldsheet model is:
\begin{equation}
    T_{B} = \frac{1}{2} \Lambda^{\alpha} \partial \Omega_{\alpha} - \frac{1}{2} \Omega_{\alpha} \partial \Lambda^{\alpha} - \frac{1}{2} \psi^{m} \partial \psi_{m} + T_{J},
\end{equation}
where $T_{J}$ is the stress-energy tensor associated to the current algebra.



The OPEs satisfied by the canonical variables are 
\begin{equation}\label{standardOPEs}
    \big\llangle \Lambda^{\alpha}_{(z)} \Omega_{\beta (w)} \big\rrangle = \frac{\delta^{\alpha}_{\beta}}{ (z-w)}, \ \ \ \  \big\llangle \psi^{m}_{(z)} \psi^{n}_{(w)} \big\rrangle = \frac{\eta^{mn}}{(z-w)},
\end{equation}
where we have introduced the notation $\llangle \rrangle$ to mean the singular terms in an OPE. Whenever there is no room for confusion we will just refer to OPE to mean that we are interested in the corresponding singular terms.

The OPEs/algebra satisfied by the constraints $(T_{B}, T_{F}, G^{\alpha})$ are readily found to be
\begin{align} \label{constraintalgebra}
      \big\llangle G^{\alpha}_{(z)}T_{F (\omega)} \big\rrangle & = - \frac{2}{(z-w)} \Lambda^{\alpha} T_{F (w)}, \\[1.0ex]
      \big\llangle T_{B (z)}G^{\alpha}_{(w)} \big\rrangle & = \frac{3}{2(z-w)^{2}} G^{\alpha}_{(w)} + \frac{1}{(z-w)} \partial G^{\alpha}_{(w)}, \\[1.0ex]
      \big\llangle T_{B(z)}T_{F(w)} \big\rrangle & = \frac{3}{2(z-w)^{2}} T_{F (w)} + \frac{1}{(z-w)} \partial T_{F(w)}, \\[1.0ex]
      \big\llangle T_{F(z)}T_{F(w)} \big\rrangle &=  0, \\[1.0ex]
      \big\llangle T_{B(z)}T_{B(w)} \big\rrangle &=  \frac{-\frac{11}{2} + \frac{c_{J}}{2}}{(z-w)^{4}} + \frac{2}{(z-w)^{2}}T_{B(w)} + \frac{1}{(z-w)}\partial T_{B(w)}, \\[1.0ex]
       \big\llangle G^{\alpha}_{(z)}G^{\beta}_{(w)} \big\rrangle &= -\frac{4}{(z-w)} \Lambda^{[\alpha}G^{\beta]}_{(w)} - \frac{56}{(z-w)^{2}} \Lambda^{\alpha}\Lambda^{\beta}_{(w)} - \frac{36}{(z-w)} \partial \Lambda^{\beta} \Lambda^{\alpha}_{(w)} \nonumber \\[1.0ex] & \hspace{-2.2cm} - \frac{20}{(z-w)}\partial \Lambda^{\alpha} \Lambda^{\beta}_{(w)} + \frac{16}{(z-w)^{2}} (\gamma^{m})^{\alpha \beta} (\Lambda \gamma_{m} \Lambda)_{(w)} + \frac{16}{(z-w)} (\gamma_{m})^{\alpha \beta} (\partial \Lambda \gamma^{m} \Lambda)_{(w)}. \label{GGconstraintalgebra}
\end{align}

As already noticed in \cite{Berkovits:2019bbx}, the above construction is not limited to the heterotic case and one can readily generalize to the Type IIB case. \\

\textbf{The BRST Operator.} Due to the reducibilities \eqref{one-reducibility} and \eqref{two-reducibility} the BRST operator for the model \eqref{heteroriginalaction} is non-trivial to find. In general, for worldline systems, when one is in presence of a set of constraints $G_{a_{0}}$, \ $a_{0} = 1, \ldots, m_{0},$ that have a set of reducibilities: $$Z_{a_{1}}^{\ a_{0}}G_{a_{0}} = 0,  \ \ \ \ \ \ a_{1}=1,\ldots,m_{1},$$ which may themselves be reducible: $$Z_{a_{k}}
^{\ a_{k-1}}Z_{a_{k-1}}^{\ a_{k-2}} = C_{a_{k}}^{\ a_{k-2}a_{0}}G_{a_{0}}, \ \ \ k=1, \ldots, L, \ \ \ a_{k} = 1, \ldots, m_{k},$$
one introduces a ghost-for-ghost pair $(\eta_{a_{k}}, \rho_{a_{k}})$ with total ghost number $(k+1,-(k+1))$ for each level of reducibility, in addition to the standard ghosts $(\eta_{a_{0}}, \rho_{a_{0}})$ associated to the constraints $G_{a_{0}}$. A general prescription \cite{Henneaux:1992ig} to write down the BRST operator is given by:
\begin{equation}
    Q = \eta^{a_{0}}G_{a_{0}} + \eta^{a_{k}}Z_{a_{k}}^{\ a_{k-1}}\rho_{a_{k-1}} + \bar{Q},
\end{equation}
where $\bar{Q}$ stands for further terms in $Q$ containing at least two $\eta$'s and one $\rho$ or two $\rho$'s and one $\eta$ and that are constructed such that $Q$ is nilpotent.

It is straightforward to adapt the previous construction to the case of a 2D CFT. We introduce, apart from the standard reparametrization ghosts $(c,b)$ and the ghosts $(g_{\alpha},f^{\alpha}), \, (\gamma, \beta)$ associated to the constraints $G^{\alpha}$ and $T_{F}$ respectively, a pair of (bosonic) ghosts-for-ghosts $(s_{m},t^{m})$ and (fermionic) ghosts-for-ghosts $(\eta, \rho)$. One then finds the BRST current to be: 
\begin{equation} \label{originalmodel-brstcurrent}
     q(z)  = \Big( cT + b c \partial c + q_{sp} -16 \Lambda^{\alpha} \partial g_{\alpha} \Big)(z),
\end{equation}
where
\begin{align} \label{originalmodel-superparticle-brstcurrent}
    q_{sp}(z) = &\bigg(g_{\alpha} G^{\alpha} + \gamma T_{F} + s_{m}(\Lambda \gamma^{m} f) + s_{m} (2 \psi^{m} \beta) + \eta (\Lambda \gamma_{m} \Lambda) t^{m}  \nonumber \\[1.0ex]
    & \ \ + 2 (\Lambda^{\alpha} g_{\alpha})(g_{\beta} f^{\beta})
    - 2 (\Lambda^{\alpha} g_{\alpha})(\gamma \beta)
    -2 \big[ \Lambda^{\alpha} (\gamma_{n})_{\alpha \beta} (\gamma^{m})^{\beta \gamma} g_{\gamma} \big] s_{m}t^{n}  \nonumber \\[1.0ex]
    & \hspace{4.2cm} + 4 (\Lambda^{\alpha} g_{\alpha})\eta \rho + 2 \eta^{nm} s_n s_m \rho - \eta \beta^{2} \bigg)(z).
\end{align}
We have separated the contributions associated to the Virasoro constraint to those associated to the superparticle terms in \cite{Sepulveda:2020kjc}. Single contraction contributions in the $q_{sp}(z) q_{sp}(w)$ OPE vanish in a tantamount computation to that of the superparticle \cite{Sepulveda:2020kjc}. Notice, however, that new contributions could in principle arise in the simple poles due to the expansion of double contraction terms which could render the BRST operator non-nilpotent. There are two of these type of contributions: those which contain two $g$-ghosts, and those that are proportional to the $s_{m}$ ghost-for-ghost. For instance, a double contraction of the $(g,f)$ ghosts in the $2 (\Lambda^{\alpha} g_{\alpha})(g_{\beta} f^{\beta})$ term with itself gives a simple pole contribution of:
\begin{equation*}
    4 [(\Lambda g)(g f)] (z) [(\Lambda g)(g f)](w) \xrightarrow[\begin{subarray}{l} \text{($g$,$f$)} \\ \text{($g$,$f$)} \end{subarray}]{\text{}} \frac{\Big[ 52 (\partial \Lambda g)(\Lambda g) + 60 (\Lambda \partial g)(\Lambda g) \Big] (w)}{(z-w)},
\end{equation*}
where the variables below the arrow stand for the fields that we have contracted, and the right side is the corresponding contribution to the simple pole. This is an example of a contribution arising from the expansion of a double contraction and that contains two $g$-ghosts. The other non-zero contributions of this type are:
\begin{align}
    & 4 (\Lambda g)(\gamma \beta)_{(z)}(\Lambda g)(\gamma \beta)_{(w)} \xrightarrow[\begin{subarray}{l} \text{($\gamma$,$\beta$)}\\ \text{($\gamma$,$\beta$)} \end{subarray}]{\text{}} -4\frac{\Big[ (\partial \Lambda g)(\Lambda g) + (\Lambda \partial g)(\Lambda g) \Big] (w)}{(z-w)}, \nonumber \\
    & 4 (\Lambda g)(\eta \rho)_{(z)}(\Lambda g)(\eta \rho)_{(w)} \xrightarrow[\begin{subarray}{l} \text{($\eta$,$\rho$)}\\ \text{($\eta$,$\rho$)} \end{subarray}]{\text{}} 16\frac{\Big[ (\partial \Lambda g)(\Lambda g) + (\Lambda \partial g)(\Lambda g)\Big] (w)}{(z-w)}, \nonumber \\
    & 4 (\Lambda \gamma_{n} \gamma^{m} g)s_{m}t^{n}_{(z)} (\Lambda \gamma_{p} \gamma^{q} g)s_{q}t^{p}_{(w)} \xrightarrow[\begin{subarray}{l} \text{($s$,$t$)} \\ \text{($s$,$t$)} \end{subarray}]{\text{}} \frac{\big[ 16(\Lambda \gamma^{m} \Lambda)(\partial g \gamma_{m} g) - 16(\partial \Lambda)(\Lambda g) - 80 (\Lambda \partial g) (\Lambda g)\big](w)}{(z-w)}. \nonumber
\end{align}
There is also a contribution from the double poles in the $g_{\alpha}G^{\alpha}(z)g_{\beta}G^{\beta}(w)$ OPE that can be read from the $G^{\alpha}G^{\beta}$ OPE and expanding $z$ around $w$:
\begin{equation*}
    \frac{16\big[ (\partial \Lambda g)(\Lambda g)\big](w)}{(z-w)} - \frac{56\big[ (\Lambda \partial g)(\Lambda g) \big](w)}{(z-w)} + \frac{16\big[ (\Lambda \gamma^{m} \Lambda)(\partial g \gamma_{m} g)\big](w)}{(z-w)}.
\end{equation*}
Adding all these type of contributions, one gets:
\begin{equation} \label{ccdoublepole}
    \frac{ \big[ 32(\Lambda \gamma^{m} \Lambda) (\partial g \gamma_{m} g) + 64(\partial \Lambda g)(\Lambda g) - 64(\Lambda \partial g)(\Lambda g)\big](w)}{(z-w)}.
\end{equation}
In order for the BRST operator to be nilpotent we must cancel these contributions. Quite remarkably, this issue is fully-handled just by the last term in \eqref{originalmodel-brstcurrent} which corresponds to a normal ordering term that has been added to consider these contributions. Specifically, the $-16\Lambda^{\alpha} \partial g_{\alpha}$ term in \eqref{originalmodel-brstcurrent} introduces further OPEs with $q_{sp}$, namely: 
\begin{align}
    & -16 \Lambda^{\alpha}\partial g_{\alpha (z)} g_{b}G^{b}_{(w)} + \textrm{cross-term} \xrightarrow[\text{($\Lambda$,$\Omega$)}]{\text{}} \frac{ \big[ -32(\Lambda \gamma^{m} \Lambda)(\partial g \gamma_{m} g) + 64 (\Lambda \partial g)(\Lambda g) \big](w)}{(z-w)}, \\
    & -32 \Lambda^{\alpha}\partial g_{\alpha (z)} (\Lambda g)(gf)_{(w)} + \textrm{cross-term} \xrightarrow[\text{($\Lambda$,$\Omega$)}]{\text{}} -\frac{64\Big[ (\partial \Lambda g)(\Lambda g)\Big](w)}{(z-w)},
\end{align}
which precisely cancel \eqref{ccdoublepole} and no terms with two $g$-ghosts remain. Furthermore, the single term $-16\Lambda^{\alpha} \partial g_{\alpha}$ also handles the second type of contributions proportional to $s_{m}$. All these contributions are given by:
\begin{align}
    & 2 \gamma (\Lambda \gamma^{m} \Lambda) \psi_{m (z)} s_{n}\psi^{n}\beta_{(w)} + \textrm{cross-term} \xrightarrow[\begin{subarray}{l} \text{($\psi$,$\psi$)} \\ \text{($\beta$,$\gamma$)} \end{subarray}]{\text{}} \frac{[4(\partial \Lambda \gamma^{m} \Lambda)s_{m} - 2(\Lambda \gamma^{m} \Lambda) \partial s_{m}](w)}{(z-w)}, \nonumber \\[1.5ex]
   - & 16\Lambda \partial g_{(z)} s_{m}(\Lambda \gamma^{m} f)_{(w)} + \textrm{cross-term} \xrightarrow[\text{($g$,$f$)}]{\text{}} -\frac{16(\Lambda \gamma^{m} \Lambda) \partial s_{m} (w)}{(z-w)}, \nonumber \\[1.5ex]
   - & 2(\Lambda \gamma_{n}\gamma^{m} g) s_{m}t^{n}_{(z)} s_{p}(\Lambda \gamma^{p} f)_{(w)} + \textrm{cross-term} \xrightarrow[\begin{subarray}{l} \text{($s$,$t$)} \\ \text{($g$,$f$)} \end{subarray}]{\text{}} -\frac{16(\Lambda \gamma^{m} \Lambda) \partial s_{m} (w)}{(z-w)}, \nonumber \\
    & g_{\alpha}G^{\alpha}_{(z)}s_{m}(\Lambda \gamma^{m} f)_{(w)} + \textrm{cross-term} \xrightarrow[\begin{subarray}{l} \text{($\Lambda$,$\Omega$)} \\ \text{($g$,$f$)} \end{subarray}]{\text{}} \frac{[14(\Lambda \gamma^{m} \Lambda) \partial s_{m} - 28(\partial \Lambda \gamma^{m} \Lambda) s_{m}](w) }{(z-w)}, \nonumber \\[1.0ex]
    & 2 \eta^{mn}s_{m}s_{n}\rho_{(z)}\eta(\Lambda \gamma^{p} \Lambda)t^{p}_{(w)} + \textrm{cross-term} \xrightarrow[\begin{subarray}{l} \text{($s$,$t$)} \\ \text{($\eta$,$\rho$)} \end{subarray}]{\text{}} \frac{[4(\Lambda \gamma^{m} \Lambda) \partial s_{m} - 8(\partial \Lambda \gamma^{m} \Lambda) s_{m}](w) }{(z-w)}. \nonumber
\end{align}
As it is easy to see, the term $-16\Lambda^{\alpha} \partial g_{\alpha}$ again takes care of all these contributions and the final result is a total derivative:
\begin{equation}
    -16 \frac{\partial[(\Lambda \gamma^{m} \Lambda)s_{m}](w)}{(z-w)},
\end{equation}
which is sufficient for the BRST operator to be nilpotent.

Clearly, we also need to consider the presence of the Virasoro constraint in the full BRST current:
\begin{equation}
    T = T_{B} + T_{\textrm{gh}},
\end{equation}
where $T_{\textrm{gh}}$ corresponds to the stress energy tensor of all ghost fields others than the $(b,c)$ system. This is straightforward to accommodate with the superparticle contributions recalling that the BRST current has conformal weight one.

\section{The Pure Spinor Twistor Ambitwistor String} \label{section3}

In this section we construct an ambitwistor worldsheet model based on the description of the $D=10$ massless superparticle developed in \cite{Sepulveda:2020kjc}. We start defining the model and considering the OPEs for many quantities of interest in analogy with the standard pure spinor formalism \cite{Berkovits:2000fe, Oda:2005sd, Oda:2007ak}. We take special care of the fact that the OPEs satisfied by our variables do not correspond to those of a free theory. After a convenient gauge fixing procedure, a simple BRST operator is constructed and physical states are defined.

\subsection{The Worldsheet Model}

The variables from which the worldsheet model is going to be defined are given by
\begin{equation}
    \mathcal{Z}^{I} = (\lambda^{\alpha},\, \mu_{\alpha},\, \Gamma^{m}), \ \ \ \ \ \bar{\mathcal{Z}}_{I} = (\barmu_{\alpha},\, -\barlambda^{\alpha}, \bar{\Gamma}^{m}),
\end{equation}
where $\lambda^{\alpha}$ is a pure spinor, $\barlambda^{\alpha}$ is a 16-component spinor, and $\Gamma^{m}$ is a fermionic vector. $\bar{\mathcal{Z}}_{I}$ will correspond to the canonical conjugates to the variables defining $\mathcal{Z}^{I}$. By definition the variables are required to solve:
\begin{align} \label{psconstraints}
S^{m} & \coloneqq \lambda\gamma^{m}\lambda = 0, \\[1.0ex]
D & \coloneqq \lambda\mu = 0, \\[1.0ex] \Phi^{mn} & \coloneqq (\lambda\gamma^{mn}\mu) + 4\Gamma^{m}\Gamma^{n} = 0, \\[1.0ex] E_{\alpha} & \coloneqq (\lambda\gamma^{m})_{\alpha}\Gamma_{m} = 0, \label{psconstraints2}
\end{align}
and are related to the standard superspace variables through the ``incidence relations'':
\begin{equation} \label{incidencerelationsps}
    \mu_{\alpha} = (\gamma_{m} \lambda)_{\alpha} X^{m} + \Gamma^{m}(\gamma_{m} \theta), \ \ \ \ \ \Gamma^{m} = (\lambda \gamma^{m} \theta).
\end{equation}
In virtue of \eqref{psconstraints}-\eqref{psconstraints2}, physical quantities must be invariant under the gauge transformations:
\begin{align}
    \delta \barmu_{\alpha} & = \mu_{\alpha} d + (\gamma^{mn} \mu)_{\alpha} \phi_{mn} + (\gamma^{m} \epsilon)_{\alpha} \Gamma_{m} + (\gamma^{m} \lambda)_{\alpha} s_{m}, \\ \delta \barlambda^{\alpha} & = -\lambda^{\alpha} d - (\lambda \gamma^{mn})^{\alpha} \phi_{mn}, \\ \delta \bar{\Gamma}_{m} & = - 8 \phi_{mn} \Gamma^{n} + (\lambda \gamma_{m} \epsilon),
\end{align}
where $s_{m}$, $d$, $\phi_{mn}$, $\epsilon_{\alpha}$ are gauge parameters associated to \eqref{psconstraints}-\eqref{psconstraints2} respectively. 

The OPEs satisfied by these variables are non-trivial considering the relations \eqref{psconstraints}-\eqref{psconstraints2} which effectively render the theory interacting. The OPEs that we have to consider are given by:
\begin{align}
\llangle \bar{\mu}_{\alpha (z)} \lambda^{\beta}_{(w)} \rrangle &=  \frac{-1}{(z-w)}\Big(\delta^{\beta}_{\alpha} - K_{\alpha}^{\ \beta} \Big)_{(w)} \, , \label{OPElambdaw} \\[0.4ex]
\llangle \bar{\lambda}^{\beta}_{(z)} \mu_{\alpha (w)} \rrangle & =  \frac{1}{(z-w)}K_{\alpha (w)}^{\ \beta}\,, \\[0.4ex]
\llangle \bar{\mu}_{\alpha (z)} \mu_{\beta (w)} \rrangle & = \frac{1}{(z-w)}\Big( Y_{\alpha}\mu_{\beta} + Y_{\beta}\mu_{\alpha} - \frac{1}{2}\gamma^{m}_{\alpha\beta}(Y \gamma_{m}\mu) \Big)_{(w)}\,, \\[0.4ex]
\llangle \bar{\mu}_{\alpha (z)} \Gamma^{m}_{(w)} \rrangle & = \frac{1}{2(z-w)}\Big((\gamma^{p} \gamma^{m} Y)_{\alpha}\Gamma_{p}\Big)_{(w)}\,, \\[0.4ex]
\llangle \bar{\Gamma}^{n}_{(z)} \Gamma^{m}_{(w)} \rrangle & = \frac{1}{2 (z-w)}(\lambda \gamma^{m}\gamma^{n} Y)_{(w)}\,, \\[0.4ex]
\llangle \bar{\Gamma}^{m}_{(z)} \mu_{\alpha (w)} \rrangle & = \frac{-1}{(z-w)}\Big( (\gamma^{p}\gamma^{m} Y)_{\alpha}\Gamma_{p}\Big)_{(w)}, \label{OPEmugammabar}
\end{align}
where we have defined the projector
\begin{equation}
    K_{\alpha}^{\ \beta} = \frac{1}{2}(\lambda \gamma_{s})_{\alpha}(\gamma^{s} Y)^{\beta},
\end{equation}
with
\begin{equation}
    Y_{\alpha} = \frac{\nu_{\alpha}}{(\lambda \nu)},
\end{equation}
and $\nu_{\alpha}$ a fixed pure spinor so that $Y \gamma^{m} Y = 0$. 

These OPEs can be found by requiring that the OPE between any single conjugate variable $\barlambda^{\alpha}$, $\barmu_{\alpha}$, or $\bar{\Gamma}^{m}$ has a vanishing OPE with the corresponding constraints \eqref{psconstraints}-\eqref{psconstraints2}. Notice that this construction is similar in fashion to that of the standard pure spinor formalism as formulated in \cite{Oda:2005sd, Oda:2007ak}. In order to deal with the OPEs \eqref{OPElambdaw}-\eqref{OPEmugammabar} we have to be careful with the fact that the coefficients are non-constant and thus we have to resort to the tools outlined in Appendix \ref{AppendixA} to proceed. In particular, we have to be careful with the ordering of the different operators when constructing the theory. As explained in Appendix \ref{AppendixA}, we define the normal-ordered product
\begin{equation} \label{NO}
    (AB)_{(w)} = \oint_{w} \frac{\mathrm{d}x}{(x-w)} A_{(x)} B_{(w)},
\end{equation}
which let us consistently separate the finite terms from the divergent ones as $z \rightarrow w$ in $A_{(z)}B_{(w)} = \llangle A_{(z)} B_{(w)}\rrangle + (A_{(z)} B_{(w)})$.

The ambitwistor worldsheet model that we are going to consider here is based on a model of the $D=10$ massless superparticle developed by the authors in a complementary paper \cite{Sepulveda:2020kjc}. The pure spinor twistor heterotic ambitwistor string action is defined as:
\begin{eqnarray}\label{psambiaction}
S &=& \int d^{2}z\,\bigg(\bar{\mathcal{Z}}_{I}\bar{\partial}\mathcal{Z}^{I} + \zeta J + \chi B + \Upsilon_{abc}\tilde{\mathcal{M}}^{abc}\bigg) + S_{J},
\end{eqnarray}
where $S_{J}$ stands for the current algebra action, and $\zeta$, $\xi$, and $\Upsilon_{abc}$ are Lagrange multipliers enforcing the (classical) constraints:
\begin{align}
J_{0} & = \bar{\mu}_{\alpha}\lambda^{\alpha} - \bar{\lambda}^{\alpha}\mu_{\alpha} + \bar{\Gamma}_{m}\Gamma^{m}, \label{defJ}\\
B & = (\lambda\gamma^{m}\bar{\lambda})\bar{\Gamma}_{m} - (\bar{\lambda}\gamma^{m}\bar{\lambda})\Gamma_{m}, \\
\tilde{\mathcal{M}}^{abc} & = (\lambda\gamma^{[a}\bar{\lambda})N^{bc]} + \frac{1}{12}(\tilde{q}\gamma^{abc}\tilde{q}), \label{defMabc}
\end{align}
where $a,b,c$ are $SU(5)$ fundamental indices. $N^{mn}$ and $\tilde{q}_{\alpha}$ are the super-Lorentz generators given by
\begin{eqnarray}
N_{0}^{mn} &=& -\frac{1}{2}(\barmu \gamma^{mn} \lambda) + \frac{1}{2}(\bar{\lambda}\gamma^{mn}\mu) - 2(\bar{\Gamma}^{[m}\Gamma^{n]}), \label{defNmn}\\
\tilde{q}_{\alpha} &=& (\gamma^{m}\bar{\lambda})_{\alpha}\Gamma_{m} - \frac{1}{2}(\gamma^{m}\lambda)_{\alpha}\bar{\Gamma}_{m}. \label{defqtilde}
\end{eqnarray}
We have written a subindex $0$ whenever we have taken a quantity without considering correction terms due to quantum-mechanical effects. 

In order to consider the corresponding corrections to the operators, we need to be careful with the corresponding ordering of the operators according to \eqref{NO} and ask for the correct OPEs to be fulfilled. For instance, when one considers the corrected Lorentz generator $N^{mn}$:
\begin{equation}
    N^{mn} = -\frac{1}{2}(\barmu \gamma^{mn} \lambda) + \frac{1}{2}(\barlambda \gamma^{mn} \mu) -2(\bar{\Gamma}^{[m}\Gamma^{n]}) - \partial \big(Y \gamma^{mn} \lambda \big),
\end{equation}
where the parenthesis stand as well for the normal ordering \eqref{NO}, one finds that the singular terms in the $N^{mn}_{(z)}N^{pq}_{(w)}$ OPE are given by:
\begin{equation}
    \big\llangle N^{mn}_{(z)} N^{pq}_{(w)} \big\rrangle = -3 \frac{(\eta^{mq} \eta^{np} - \eta^{mp} \eta^{nq})}{(z-w)^{2}}+ \frac{N^{mp} \eta^{nq} - N^{np} \eta^{mq} + N^{nq} \eta^{mp} - N^{mq} \eta^{np}}{(z-w)},
\end{equation}
which means that $N^{mn}$ generates a level $k=-3$ current algebra. The total derivative is responsible of cancelling out all spurious terms that would make the current $N^{mn}$ not satisfy a Kac-Moody algebra (see appendix \ref{AppendixA} for further details).

We can perform a similar analysis for the projective weight operator $J$ and the stress energy tensor $T$. Considering correction terms and using the OPEs \eqref{OPElambdaw}-\eqref{OPEmugammabar} we find that: 
\begin{align}
J & =  \bar{\mu}_{\alpha}\lambda^{\alpha} - \mu_{\alpha}\bar{\lambda}^{\alpha} + \bar{\Gamma}_{m}\Gamma^{m} - 4(Y_{\alpha}\partial\lambda^{\alpha}), \\[1.0ex]
T & = -\bar{\mu}_{\alpha}\partial\lambda^{\alpha} + \bar{\lambda}^{\alpha}\partial \mu_{\alpha} - \bar{\Gamma}_{m}\partial\Gamma^{m} + 2\partial(Y_{\alpha}\partial\lambda^{\alpha}),
\end{align}
with the full set of non-trivial OPEs given by:
\begin{align}
    & \llangle T_{(z)}T_{(w)} \rrangle = \frac{22+ c_{J}}{{2 (z-w)^{4}}} + \frac{2T_{(w)}}{(z-w)^{2}} + \frac{\partial T_{(w)}}{(z-w)}, \label{TTope}\\[1.0ex]
    & \llangle J_{(z)} J_{(w)} \rrangle = - \frac{3}{(z-w)^{2}}, \label{JJope}\\[1.0ex]  & \llangle T_{(z)}J_{(w)} \rrangle = -\frac{7}{(z-w)^{3}} + \frac{J_{(w)}}{(z-w)^{2}} + \frac{\partial J_{(w)}}{(z-w)}, \label{TJope}\\[1.0ex]
    & \llangle T_{(z)}N^{mn}_{(w)} \rrangle = \frac{N^{mn}_{(w)}}{(z-w)^{2}} + \frac{\partial N^{mn}_{(w)}}{(z-w)}. \label{TNope}
\end{align}
In particular, the Lorentz generator $N^{mn}$ behaves properly as a primary field, $J$ has a regular OPE with $N^{mn}$, and $J$ exhibits a conformal anomaly of $-7$. Interestingly, the quantum corrections can be arranged in a unique way such that the Lorentz covariance of the OPEs \eqref{TTope}-\eqref{TNope} is preserved. That is, $Y$ spinors appear in the OPEs only in the quantum corrections to $J$, $N^{mn}$ and $T$. Compare this, for instance, to the OPEs of the variables defining the model \eqref{OPElambdaw}-\eqref{OPEmugammabar}. 




The components of the constraint $\tilde{\mathcal{M}}^{abc}$ in \eqref{defMabc} are not all independent. Indeed, careful inspection of $\tilde{\mathcal{M}}^{abc}$ leads us to the conclusion that it satisfies the following reducibility relations:
\begin{align}
H^{abcd} & \coloneqq P^{[a}\tilde{\mathcal{M}}^{bcd]} = -\frac{1}{24}\epsilon^{abcde}\tilde{q}_{e}\lambda^{+}B\label{red1},\\[1.0ex]
H^{abcde} & \coloneqq P^{[a}H^{bcde]} = 0\label{red2}.
\end{align}
As done for $J$, $N^{mn}$, and $T$, one can also compute the quantum corrections for $\tilde{\mathcal{M}}^{abc}$ by requiring that it behaves as a rank three tensor under Lorentz transformations. However, since this will not be relevant for our study and it is always possible to choose a frame where the quantum correction vanishes, we will ignore this ambiguity.

The central charge can be calculated for the whole system in the usual way. Using that ($J$, $B$, $\tilde{\mathcal{M}}^{abc}$) are conformal weight $(1, 2, 2)$ operators, together with eqns. \eqref{red1}, \eqref{red2}, one obtains:
\begin{equation}
c_{het.} = 22 + 10 - 10 - 26 - 2 + 26 - 10\times 26 + 5\times 74 - 1\times 146 + c_{J}.
\end{equation}
As in the $SO(32)$ or $E_{8}\times E_{8}$ heterotic strings, one finds that the total central charge vanishes when $c_{J} = 16$.

It is straightforward to generalize the above construction for the Type IIB case. The action is given by
\begin{equation} \label{psambiactiontypeIIB}
S = \int d^{2}z\,\bigg(\bar{\hat{\mathcal{Z}}}_{I}\bar{\partial}\hat{\mathcal{Z}}^{I} + \zeta J + \chi B + \hat{\chi}\hat{B} +  \Upsilon_{abc}\mathcal{\tilde{M}}^{abc}\bigg),
\end{equation}
where the supertwistor variables $\hat{\mathcal{Z}}^{I} = (\lambda^{\alpha}, \mu_{\alpha}, \Gamma^{m}, \hat{\Gamma}^{m})$, $\bar{\hat{\mathcal{Z}}}_{I} = (\bar{\mu}_{\alpha}, -\bar{\lambda}^{\alpha}, \bar{\Gamma}^{m}, \bar{\hat{\Gamma}}^{m})$ are defined to satisfy the analogs of \eqref{psconstraints}-\eqref{psconstraints2}:
\begin{align}
& \lambda\gamma^{m}\lambda = 0, \ \ \ \lambda\mu = 0, \ \ \ (\lambda\gamma^{mn}\mu) + 4\Gamma^{m}\Gamma^{n} = 0, \\[1.0ex] & (\lambda\gamma^{m})_{\alpha}\Gamma_{m} = 0, \hspace{1.3cm} (\lambda\gamma^{m})_{\alpha}\Gamma_{m} = 0.
\end{align}
The constraints $\hat{B}$, $\tilde{M}^{abc}$ are defined by
\begin{eqnarray}
\tilde{B} &=& (\lambda\gamma^{m}\bar{\lambda})\bar{\tilde{\Gamma}}_{m} - (\bar{\lambda}\gamma^{m}\bar{\lambda})\tilde{\Gamma}_{m}, \\[1.0ex]
\tilde{M}^{abc} &=& (\lambda\gamma^{[a}\bar{\lambda})N^{bc]} - \frac{1}{12}(\tilde{q}\gamma^{abc}\tilde{q}) - \frac{1}{12}(\hat{\tilde{q}}\gamma^{abc}\hat{\tilde{q}}), \label{mtildetypeIIB}
\end{eqnarray}
where $\hat{\tilde{q}}_{\alpha}$, $\tilde{q}_{\alpha}$, $N^{mn}$ are the type IIB super-Lorentz generators. As before, one can show that the constraint $\tilde{M}^{abc}$ satisfies reducibility relations similar to those in eqns. \eqref{red1}, \eqref{red2}.  
The total central charge is then easily shown to vanish:
\begin{equation}
c_{IIB} = 22 + 10 - 10 - 10 - 26 - 2 + 26 + 26 - 10\times 26 + 5\times 74 - 1\times 146 = 0 .
\end{equation}

\subsection{Construction of Physical States}
Physical states are defined via a BRST operator that will be constructed by applying the Faddeev-Popov method for the constraint $B$ and keeping the volume of the remaining symmetry groups as factors dividing out the path integral measure. In this manner, the BRST current reads
\begin{equation}\label{brstb}
q{(z)} = \gamma B{(z)} = (\eta e^{\phi} B){(z)},
\end{equation}
where $(\beta,\gamma)$ are the ghosts for the symmetry $B$, which have been fermionized through the standard procedure:
\begin{equation}
\gamma = \eta e^{\phi}, \ \ \ \beta = \partial \xi e^{-\phi},
\end{equation}
and $(\eta,\xi)$, $\phi$ are $bc$-type, linear dilaton CFTs, respectively. These definitions give rise to the picture charge defined by
\begin{equation}
N_{p} = \int dz\, (\xi\eta - \partial\phi).
\end{equation}

In order to write down vertex operators we will introduce additional $SU(5)$ indices on the pure spinor twistor variables which accounts for the symmetry group under which the twistor-like constraint $(\lambda \gamma^{m})_{\alpha}P^{m}=0$ is left invariant. This construction is essentially similar to the ones developed in \cite{Cheung:2009dc,Boels:2012ie,Geyer:2018xgb,Geyer:2019ayz} where spinor helicity variables describing massless states are assigned an additional little group index. Hence, the twistor variables now take the form $(\lambda_{r\,\dot{a}}^{\alpha}, \bar{\pi}_{r}^{\beta\,\dot{b}})$, where $\dot{a}$, $\dot{b}$ are $SU(5)$ vector indices, and satisfy
\begin{equation} \label{lambda-pibar-andk}
    \lambda_{r\,\dot{a}}^{\alpha}(\gamma^{m})_{\alpha\beta}\bar{\pi}_{r}^{\beta\,\dot{b}} = k_{r}^{m}\delta^{\dot{b}}_{\dot{a}}.
\end{equation}
Using these variables one can define the following ten-dimensional spinors:
\begin{equation}
\lambda^{\alpha}_{r} \coloneqq \lambda^{\alpha}_{r\,\dot{a}}\epsilon^{\dot{a}}_{r}, \ \ \ \hspace{2mm}\bar{\pi}^{\alpha}_{r} \coloneqq \bar{\pi}_{r}^{\alpha\,\dot{a}}\bar{\epsilon}_{r\,\dot{a}},
\end{equation}
where $\epsilon^{\dot{a}}_{r}$, $\bar{\epsilon}_{r\,\dot{a}}$ are $SU(5)$ vectors. Notice that these spinors satisfy the relation
\begin{equation}
\lambda^{\alpha}_{r}(\gamma^{m})_{\alpha\beta}\bar{\pi}_{r}^{\beta} = k_{r}^{m}(\epsilon^{\dot{a}}_{r}\bar{\epsilon}_{r\,\dot{a}}),
\end{equation}
so in order to recover the more standard twistor relation $\lambda^{\alpha}_{r}(\gamma^{m})_{\alpha\beta}\bar{\pi}^{\beta}_{r} = k^{m}_{r}$ we integrate the $\epsilon_{r}$- and $\bar{\epsilon}_{r}$-variables in the vertex operator expression along with a delta function ensuring an appropriate localization. The super-Yang-Mills vertex operator at picture -1 then reads
\begin{align}\label{vertexoperator}
U^{(-1)}_{r} & = \int d^{2}z_{r}\,d^{5}\epsilon_{r}\,d^{5}\bar{\epsilon}_{r}\, \bar{\delta}(\epsilon_{r}^{\dot{a}}\bar{\epsilon}_{r\,\dot{a}} - 1)\bar{\delta}^{10}\bigg(\frac{\lambda_{ab}}{\lambda^{+}}(z_{r}) - \frac{\lambda_{ab\,\dot{a},r}\epsilon_{r}^{\dot{a}}}{\lambda^{+}_{r\,\dot{a}}\epsilon^{\dot{a}}_{r}}\bigg) \,\phi^{\epsilon}_{r,K}(\mathcal{Z})e^{-\phi}J^{K}\nonumber\\
&= \int d^{2}z_{r}\,\Sigma^{\epsilon}_{r}\,\phi^{\epsilon}_{r,K}(\mathcal{Z})e^{-\phi}J^{K},
\end{align}
where $\Sigma^{\epsilon}_{r}$ is defined to be
\begin{equation}\label{deltafunction}
\Sigma^{\epsilon}_{r} = \int d^{5}\epsilon_{r}\,d^{5}\bar{\epsilon}_{r}\,\bar{\delta}(\epsilon_{r}^{\dot{a}}\bar{\epsilon}_{r\,\dot{a}} - 1)\bar{\delta}^{10}\bigg(\frac{\lambda_{ab}}{\lambda^{+}}(z_{r}) - \frac{\lambda_{ab\,\dot{a},r}\epsilon_{r}^{\dot{a}}}{\lambda_{r\,\dot{a}}^{+}\epsilon_{r}^{\dot{a}}}\bigg),
\end{equation}
$\lambda_{ab}$, $\lambda^{+}$ are respectively the fully antisymmetric ten-dimensional, scalar $SU(5)$ components of $\lambda^{\alpha}$, the superscript $(-1)$ stands for the picture charge, the $J^{K}$ furnish a current algebra, and $\phi^{\epsilon}_{r,K}(\mathcal{Z})$ is given by
\begin{align}
& \phi^{\epsilon}_{r,K}(\mathcal{Z}) = \bigg( \bar{s}_{r,K} + 2\Gamma_{m}a_{-\,r,K}^{m} - 4\Gamma_{m}\Gamma_{n}s_{r,K}^{mn}  \nonumber \\
& +\frac{1}{12} (\bar{\pi}_{r}\gamma_{mnpqr}\bar{\pi}_{r})\Gamma^{m}\Gamma^{n}\Gamma^{p}h^{q}a^{r}_{+\,r,K} - \frac{1}{24}  (\bar{\pi}_{r}\gamma_{mnpqr}\bar{\pi}_{r})\Gamma^{m}\Gamma^{n}\Gamma^{p}\Gamma^{q}h^{r}s_{r,K}\bigg)e^{\mu_{a}\bar{\pi}_{r}^{a}\frac{\lambda^{+}_{r}}{\lambda^{+}}},\label{twistorsuperfield}
\end{align}
which coincides with the pure spinor twistor superparticle wavefunction of \cite{Sepulveda:2020kjc, Berkovits:2009by} on the support of the delta functions of \eqref{deltafunction}. In \eqref{twistorsuperfield} the gluon polarization is given by $a_{K}^{m} = a_{-\,K}^{m} + a_{+\,K}^{m}$, with $(\bar{\pi}\gamma_{m})_{\alpha}a_{-\,K}^{m} = 0$, $(\lambda\gamma_{m})a^{m}_{+\,K} = 0$, and the gluino polarization has been split into the form $\chi^{\alpha}_{K} = \bar{\pi}^{\alpha}\bar{s}_{K} + (\gamma_{mn}\lambda)^{\alpha}s^{mn}_{K} + \lambda^{\alpha}s_{K}$, with $(\bar{\pi}\gamma_{m})_{\alpha}s^{mn}_{K} = 0$. The vector $h
^{m}$ is constant and satisfies $h^{m}(\lambda\gamma_{m}\bar{\pi}) = 1$. Moreover, we have set eleven components of $\bar{\pi}^{\alpha}$ to zero using the fact that $\lambda^{\alpha}$ is a pure spinor, so that one is left with only five components, namely $\bar{\pi}^{a}$, which transform in the fundamental of $SU(5)$. One can then consider $\bar{\pi}^{\alpha}$ to be a pure spinor, and so \eqref{twistorsuperfield} is independent of the choice of $h^{m}$.

As a check, notice that one can recover the standard exponential bosonic  contribution $e^{k_{r} \cdot X}$ to the vertex operator after using the incidence relations \eqref{incidencerelationsps} and the ten-dimensional delta function in \eqref{vertexoperator}. Indeed, one finds: $\mu_{a}\bar{\pi}^{a \, \dot{a}}_{r} \bar{\epsilon}_{r \, \dot{a}}\lambda^{+}_{r\,\dot{a}}\epsilon^{\dot{a}}_{r}/\lambda^{+} = X^{b} \lambda_{ab\,\dot{a},r}\epsilon_{r}^{\dot{a}}\bar{\pi}^{a \, \dot{b}}_{r} \bar{\epsilon}_{r \, \dot{b}} + \lambda^{+}_{r\,\dot{a}}\epsilon^{\dot{a}}_{r}\bar{\pi}_{r}^{b\,\cdot{b}}\bar{\epsilon}_{\dot{b}}X_{b}= k_{r\, a}X^{a} \epsilon_{r}^{\dot{a}} \bar{\epsilon}_{r \, \dot{a}} + k^{r\,b}X_{b}\epsilon^{\dot{a}}_{r}\bar{\epsilon}_{\dot{a}} = k_{r} \cdot X$, where we used \eqref{lambda-pibar-andk} in the second equality, and that the exponential appears on the support of the single $\bar{\delta}(\epsilon_{r}^{\dot{a}}\bar{\epsilon}_{r\,\dot{a}} - 1)$ delta function in \eqref{vertexoperator} to write down the last equality. As we will see in the next section, the delta function \eqref{deltafunction} will also give rise to the correct ten-dimensional momentum conservation delta function and the standard CHY scattering equations, thus providing further evidence on the validity of our proposal \eqref{vertexoperator}.

In order to construct vertex operators in different pictures, we define a picture-raising operator in the usual way:
\begin{equation}
Z \coloneqq \{Q,\xi\} = e^{\phi}B.
\end{equation}
We will be particularly interested in the picture number zero vertex to discuss scattering amplitudes. This is easily calculated to be
\begin{align} 
U^{(0)}_{r}  & = \int d^{2}z_{r}\, \Sigma^{\epsilon}_{r} \, \bigg[\lim_{w\rightarrow z_{r}}Z(w) \phi_{r,K}^{\epsilon}(\mathcal{Z}(z_{r}))e^{-\phi}J^{K}\bigg]  = 
\int d^{2}z_{r}\,\Sigma^{\epsilon}_{r} \,B_{-1}(\phi_{r,K}^{\epsilon}(\mathcal{Z}))J^{K}.  \label{picturenumberzerotwistorvertex}
\end{align}
The bosonic and fermionic sectors of the picture number zero vertex operator then read
\begin{eqnarray}\label{gluonpicturezero}
U^{(0)}_{bos.\,r} &=& \int\,d^{2}z_{r}\, \Sigma^{\epsilon}_{r}\,\bigg(2(\lambda\gamma_{m}\bar{\lambda})a^{m}_{-\,r,K}\nonumber\\
&& + 2M_{mn}(\lambda_{r}\gamma^{m}\bar{\pi}_{r})a^{n}_{-\,r,K} + \frac{1}{4}(\bar{\pi}_{r}\gamma_{mnpqs}\bar{\pi}_{r})(\lambda\gamma^{m}\bar{\lambda})\Gamma^{n}\Gamma^{p}h^{q}a^{s}_{+\,r,K} \nonumber\\
&& + \frac{1}{12}(\bar{\pi}_{r}\gamma_{npqst}\bar{\pi}_{r})\bar{\Gamma}^{m}\Gamma^{n}\Gamma^{p}\Gamma^{q}h^{s}(\lambda_{r}\gamma_{m}\bar{\pi}_{r})a^{t}_{+\,r,K}\bigg)J^{K}e^{\mu_{a}\bar{\pi}^{a \, \dot{a}}_{r} \bar{\epsilon}_{r \, \dot{a}} \lambda^{+}_{r\,\dot{a}}\epsilon^{\dot{a}}_{r}/\lambda^{+}}, \nonumber \\[1.0ex]
U^{(0)}_{fer.\,r} &=& \int\,d^{2}z_{r}\, \Sigma^{\epsilon}_{r} \, \bigg(\bar{\Gamma}^{m}(\lambda_{r}\gamma_{m}\bar{\pi}_{r})\bar{s}_{r,K}\nonumber\\&& - 8(\lambda\gamma^{m}\bar{\lambda})\Gamma^{n}s_{mn,K}- 4\bar{\Gamma}^{m}\Gamma_{n}\Gamma_{p}(\lambda_{r}\gamma_{m}\bar{\pi}_{r})s^{np}_{r,K} - \frac{1}{6}(\bar{\pi}_{r}\gamma_{mnpqs}\bar{\pi}_{r})(\lambda\gamma^{m}\bar{\lambda})\Gamma^{n}\Gamma^{p}\Gamma^{q}h^{s}s_{r,K}\nonumber\\
&& - \frac{1}{24}(\bar{\pi}_{r}\gamma_{npqst}\bar{\pi}_{r})\bar{\Gamma}^{m}\Gamma^{n}\Gamma^{p}\Gamma^{q}\Gamma^{s}h^{t}(\lambda_{r}\gamma_{m}\bar{\pi}_{r})s_{t,K}\bigg)e^{\mu_{a}\bar{\pi}^{a \, \dot{a}}_{r} \bar{\epsilon}_{r \, \dot{a}}\lambda^{+}_{r\,\dot{a}}\epsilon^{\dot{a}}_{r}/\lambda^{+}}. \label{picturenumberzerooperators}
\end{eqnarray}
Here, $M^{mn} \coloneqq 2\bar{\Gamma}^{[m}\Gamma^{n]}$, satisfies the same Kac-Moody algebra inside correlation functions as the RNS fermionic Lorentz currents; namely,
\begin{equation}
\big \llangle M^{mn}_{(z)}M^{pq}_{(w)} \big \rrangle= \frac{\eta^{mp}\eta^{nq} - \eta^{mq}\eta^{np}}{(z-w)^{2}} + \frac{\eta^{mp}M^{nq} - \eta^{mq}M^{np} + \eta^{nq}M^{mp} - \eta^{np}M^{mq}}{(z-w)}. \label{KacMoodyforMmn}
\end{equation}
This follows from the presence of the delta functions \eqref{deltafunction} and careful manipulations using the techniques outlined in Appendix \ref{AppendixA}.

\section{Scattering Amplitudes} \label{section4}
The tree level $N$-point correlation function will be defined to be
\begin{eqnarray}\label{amplitudeprescription}
\mathcal{A}_{N} &=& \int \frac{d^{3}\gamma_{0}\,d^{11}\lambda\,d^{5}\mu\,d^{5}\Gamma}{SL(2,\mathbb{C})\times GL(1)\times \mathbb{M}}\,\prod_{i=1}^{3} U_{i}^{(-1)} \prod_{j=4}^{N} U_{j}^{(0)},
\end{eqnarray}
where the groups $SL(2,\mathbb{C})$, $GL(1)$, $\mathbb{M}$ correspond to the symmetry groups whose generators are given by $T$, $J$, $\tilde{\mathcal{M}}^{abc}$, respectively. As is well-known, the number of killing vectors of $SL(2,\mathbb{C})$ and $GL(1)$ are three and one, respectively. On the other hand, the number of generators for $\mathbb{M}$ will be calculated here from the effective counting of zero modes for the system $\tilde{\mathcal{M}}^{abc}$ considering its reducibilities\footnote{Although we do not provide a rigorous proof for this statement, it seems to follow from a natural definition of the path integral measure in the presence of reducible symmetries. More explicitly, if a symmetry group $G$ is reducible under a subgroup $H$, we interpret its effective action into the path integral as: $<\ldots> = \int \frac{1}{\frac{Vol [G]}{Vol [H]}}\ldots$.}. Thus, one finds that $\mathbb{M}$ has $10\cdot 3 - 5\cdot 5 + 1\cdot 7 = 12$ killing vectors, which as we will see is exactly the number of zero modes needed to get the correct ten-dimensional momentum conservation delta function. Furthermore, the presence of three vertices in picture number -1 and the rest in picture number zero adequately saturate the three zero modes of the bosonic ghost $\gamma$. The measure associated to the twistor variable $\lambda^{\alpha}$ is the same as the one appearing in ordinary pure spinor strings \cite{Berkovits:2004px}, while for $\Gamma^{m}$ one has
\begin{eqnarray}\label{measureforGamma}
\int d^{5}\Gamma &=& \frac{1}{5!}(\lambda\gamma^{mnpqr}\lambda)\frac{\partial}{\partial\Gamma^{m}}\frac{\partial}{\partial\Gamma^{n}}\frac{\partial}{\partial\Gamma^{p}}\frac{\partial}{\partial\Gamma^{q}}\frac{\partial}{\partial\Gamma^{r}},
\end{eqnarray}
which respects the constraints \eqref{psconstraints2}. 

One can also view $\lambda^{\alpha}$ together with the $GL(1)$ symmetry as being a projective pure spinor. The integration measure for a projective pure spinor variable has been studied in \cite{Berkovits:2004bw}, and for the $D=10$ case reads
\begin{equation}\label{10dprojectivemeasure}
[d^{10}\lambda] = \frac{\epsilon_{\alpha_{1}\ldots\alpha_{16}}}{(\lambda^{\alpha}C_{\alpha})^{3}}d\lambda^{\alpha_{1}}\wedge\ldots\wedge d\lambda^{\alpha_{10}}\lambda^{a_{11}}(\gamma^{m}C)^{\alpha_{12}}(\gamma^{n}C)^{\alpha_{13}}(\gamma^{p}C)^{\alpha_{14}}(\gamma_{mnp})^{\alpha_{15}\alpha_{16}},
\end{equation}
where $C_{\alpha}$ is a constant spinor. Using the fact that $\lambda^{\alpha}$ is a pure spinor, one readily shows that the projective measure \eqref{10dprojectivemeasure} is independent of the choice of $C_{\alpha}$, and therefore is Lorentz-invariant.

Let us now see how the momentum conservation delta function emerges from \eqref{amplitudeprescription}. Using \eqref{lambda-pibar-andk} and the gauge described below \eqref{twistorsuperfield}, we can write
\begin{equation} \label{su5momenta}
    \lambda^{+}_{r \, \dot{a}} \epsilon^{\dot{a}}_{r} \bar{\pi}^{a \, \dot{b}}_{r} \bar{\epsilon}_{r \, \dot{b}} = k_{r}^{a} \epsilon^{\dot{a}}_{r} \bar{\epsilon}_{r \, \dot{a}}, \ \ \  \lambda_{r \, ab \, \dot{a}} \epsilon^{\dot{a}}_{r} \bar{\pi}^{b \, \dot{b}}_{r} \bar{\epsilon}_{r \, \dot{b}} = k_{r \, a} \epsilon^{\dot{a}}_{r} \bar{\epsilon}_{r \, \dot{a}},
\end{equation}
which means that the left hand sides can be replaced by $k^{a}_{r}$ or $k_{r \, a}$ correspondingly, on the support of the single $\bar{\delta}(\epsilon_{r}^{\dot{a}}\bar{\epsilon}_{r\,\dot{a}} - 1)$ delta function in \eqref{deltafunction}. This is useful after integrating out the zero modes associated to the $\mu_{a}$ field, where one is left with
\begin{equation}
    \delta^{(5)}\bigg(\sum_{r} \bar{\pi}^{a \, \dot{a}}_{r} \bar{\epsilon}_{r \, \dot{a}}\lambda^{+}_{r\,\dot{b}}\epsilon^{\dot{b}}_{r}/\lambda^{+} \bigg) = \lambda^{+} \delta^{(5)} \bigg( \sum_{r} \bar{\pi}^{a \, \dot{a}}_{r} \bar{\epsilon}_{r \, \dot{a}}\lambda^{+}_{r\,\dot{b}}\epsilon^{\dot{b}}_{r} \bigg) = \lambda^{+} \delta^{(5)}\bigg(\sum_{r} k^{a}_{r}\bigg),
\end{equation}
where we used the single delta function in \eqref{deltafunction} to write the last equality. Thus, we see that we have recovered five of the ten momentum conservation delta functions. The remaining five conditions follow from the observation that one has $N$ + 10 integration variables, $12 + 3 = 15$ killing vectors associated with $\mathbb{M}$ and $SL(2,\mathbb{C})$, and $N$ delta functions in \eqref{amplitudeprescription}, which means one is left with $N -(N + 10 - 15) = 5$ delta functions. These are exactly the five delta functions imposing the remaining momentum conservation conditions.


The integration over the non-zero modes of $\mu_{a}$ gives rise to the standard scattering equations. This follows from the standard procedure of taking the exponential in the vertex operators into the action as sources for $\barlambda^{a}$. After setting $\lambda^{+} = 1$, one finds that:
\begin{equation}
    \barlambda^{a}(z) = \sum_{r=1}^{N} \frac{\bar{\pi}^{a \, \dot{a}}_{r} \bar{\epsilon}_{r \, \dot{a}}\lambda^{+}_{r\,\dot{b}}\epsilon^{\dot{b}}_{r}}{(z-z_{r})}.
\end{equation}
Using this expression, one can compute the momentum $P^{m}(z) = (P^{a}(z), P_{a}(z))$. These $SU(5)$ components read
\begin{equation}
P^{a}(z) = \sum_{r=1}^{N}\frac{k_{r}^{a}}{z-z_{r}}, \ \ \ P_{a}(z) = \sum_{r=1}^{N}\frac{\lambda_{ab}(z_{r})\bar{\pi}_{r}^{b \dot{a}}\bar{\epsilon}_{r\,\dot{a}}\lambda^{+}_{r\,\dot{b}}\epsilon^{\dot{b}}_{r}}{z-z_{r}},
\end{equation}
where for $P^{a}(z)$ the first of \eqref{su5momenta} was used and that one is on the support of \eqref{deltafunction}. Using the ten-dimensional delta in \eqref{deltafunction}, one can work out $P_{a}(z)$:
\begin{eqnarray}
P^{a}(z) = \sum_{r=1}^{N}\frac{k_{r}^{a}}{z-z_{r}}, \ \ \ P_{a}(z) = \sum_{r=1}^{N}\frac{\lambda_{r\,ab\,\dot{c}}\epsilon_{r}^{\dot{c}}\bar{\pi}^{b\dot{b}}\bar{\epsilon}_{r\,\dot{b}} }{z-z_{r}} = \sum_{r=1}^{N}\frac{k_{r\,a}}{z-z_{r}},
\end{eqnarray}
where the second of \eqref{su5momenta} was used along with the single delta in \eqref{deltafunction}. Thus, one concludes that the delta function \eqref{deltafunction} is proportional to the standard CHY delta function $\bar{\delta}(k_{r}\cdot P(z_{r}))$.



We now move on to discuss the dependence of \eqref{amplitudeprescription} on the external polarization and momentum data. The simplest case, the 3-point function, has already been shown to correctly reproduce the standard super-Yang-Mills 3-point function in \cite{Berkovits:2009by}. We then focus here on the general case. To see that the amplitudes prescription \eqref{amplitudeprescription} indeed describes $D=10$ super-Yang-Mills interactions, we notice a close relationship between the ambitwistor string constructed in this work and the infinite tension limit of ordinary pure spinor superstrings \cite{Berkovits:2013xba}. This relation can be seen as follows: using the incidence relations \eqref{incidencerelationsps} and the constraints \eqref{psconstraints}, we see that the measure \eqref{measureforGamma} is nothing but the ordinary pure spinor measure 
\begin{equation} \label{standardpurespinormeasure}
    \langle(\lambda\gamma^{m}\theta)(\lambda\gamma^{n}\theta)(\lambda\gamma^{p}\theta)(\theta\gamma_{mnp}\theta)\rangle = 1,
\end{equation} 
up to some proportionality factor. The current algebra systems will certainly provide the same contributions in both models, so we concentrate on the sector containing the twistor superfield. It is straightforward to show that $\phi_{r,K}(Z) = V(\lambda,\theta)\,e^{k \cdot X}$ on the support of the incidence relations and the delta function \eqref{deltafunction}, where $V(\lambda,\theta) = \lambda^{\alpha}A_{\alpha}(\theta)$ is the usual pure spinor unintegrated vertex operator. Under analogous statements, the picture number zero twistor vertex is the same as the pure spinor integrated vertex operator $U(x,\theta) = [P^{m}A_{m}(\theta) + d_{\alpha}W^{\alpha}(\theta) + \frac{1}{2}N^{mn}F_{mn}(\theta)]e^{k \cdot X}$. This is essentially a direct consequence of the fact that $B$ can be shown to be proportional to the pure spinor $b$-ghost, and so the picture raising operation \eqref{picturenumberzerotwistorvertex} is nothing but the standard relation between the integrated and unintegrated vertex operators $U = \{b, V\}$. Thus, we conclude that both correlators must give the same dependence on external momentum and polarization data. 

As a check, let us consider the fully gluonic correlator. The potential difference between the present model and the ordinary pure spinor formalism lies in the picture number zero vertex \eqref{picturenumberzerotwistorvertex} and the standard integrated vertex operator. The latter has the $\theta$-expansion
\begin{equation}\label{psintegratedvertex}
U(x,\theta) = \bigg[P^{m}a_{m} + \frac{1}{2}\bigg(p\gamma^{mn}\theta + \lambda\gamma^{mn}w\bigg)k_{m}a_{n} + \ldots\bigg]e^{k \cdot X},
\end{equation}
where $\ldots$ means higher-order terms in $\theta^{\alpha}$ which can be ignored as far as scattering amplitudes is concerned by $(p_{\alpha},\theta^{\beta})$ charge conservation. The only difference between the picture number zero twistor vertex \eqref{picturenumberzerooperators} and \eqref{psintegratedvertex} is the Lorentz current inside the parenthesis in \eqref{psintegratedvertex}, which forms a level 1 Kac-Moody algebra. However, the current $M^{mn}$ in \eqref{picturenumberzerooperators} is also a level 1 Kac-Moody current algebra system as discussed below eqn. \eqref{KacMoodyforMmn}. Higher-order terms in $\Gamma^{m}$ in \eqref{picturenumberzerotwistorvertex} can also be ignored because of $(\Gamma^{m}, \bar{\Gamma}_{n})$ charge conservation, or alternatively, by using supersymmetry arguments, and so the OPEs of both models are identical. One then concludes that both correlators are equivalent to each other.

\section{Discussions and Future Directions} \label{section5}

In this work, a new description of ambitwistor strings which makes use of a set of variables and a set of reducible constraints first introduced in \cite{Berkovits:2009by, Sepulveda:2020kjc} has been presented. A detailed quantum-mechanical analysis was performed, and after introducing the ghost system associated to the fermionic symmetry $B$, a simple BRST operator was constructed. At this stage, one might wonder why one just fixes one of the gauge symmetries and leaves all the others unfixed. It turns out that if one does so, the BRST operator takes a similar form as the one found in \cite{Sepulveda:2020kjc} in the context of the superparticle. Explicitly:
\begin{eqnarray}\label{psQ}
Q &=& c T_{\textrm{full}} + \sigma J + \gamma B + f_{abc}\tilde{\mathcal{M}}^{abc} + \sigma \gamma\beta + s_{abcd} \Big[ \tilde{f}^{abc}P^{d} + \frac{1}{4!}\epsilon^{abcde}\tilde{q}_{e}\lambda^{+}\beta \Big] \nonumber \\[0.5ex]
&& + t_{abcde}P^{a}\tilde{s}^{bcde} + \frac{1}{5!}(\lambda^{+})^{2}\epsilon^{abcde}t_{abcde}\beta^{2},
\end{eqnarray}
where ghost-for-ghost fields have been introduced for each reducibility that the set of constraints satisfies. Thus the zeroth generation of constraints $J$, $B$, $T$, $\tilde{\mathcal{M}}^{abc}$ requires the introduction of the ghosts $(\sigma, \tilde{\sigma})$, $(\beta, \gamma)$, $(b,c)$, $(f_{abc}, \tilde{f}^{abc})$, respectively. The reducibility relations $\eqref{red1}$, \eqref{red2} in turn imply the presence of the ghosts-for-ghosts $(s_{abcd}, \tilde{s}^{abcd})$, $(t_{abcde},\tilde{t}^{abcde})$. Moreover, $T_{\textrm{full}}$ in \eqref{psQ} means the full stress-energy tensor
\begin{eqnarray}
T_{\textrm{full}} &=& T + \frac{1}{2}T_{bc} + T_{\beta\gamma} + T_{\sigma\tilde{\sigma}} + T_{f_{3}\tilde{f}_{3}} + T_{s_{4}\tilde{s}_{4}} + T_{t_{5}\tilde{t}_{5}},
\end{eqnarray}
with
\begin{align}
&T_{bc} = -\partial b c - 2 b\partial c, \hspace{3.3cm} T_{\beta\gamma} = -\partial\beta \gamma - 2\beta\partial\gamma, \nonumber \\
&T_{\sigma\tilde{\sigma}} = - \tilde{\sigma}\partial\sigma, \hspace{4.3cm}
T_{f_{3}\tilde{f}_{3}} = -\partial f^{abc} f_{abc} - 2 \tilde{f}^{abc}\partial f_{abc}, \nonumber\\
&T_{s_{4}\tilde{s}_{4}} = -2\partial \tilde{s}^{abcd} s_{abcd} - 3\tilde{s}^{abcd}\partial s_{abcd}, \ \ \ T_{t_{5}\tilde{t}_{5}} = -3\partial\tilde{t}^{abcde} t_{abcde} - 4\tilde{t}^{abcde}\partial t_{abcde}.
\end{align}
The BRST operator thus constructed is nilpotent up to terms proportional to $\sigma \partial\sigma$ and $\sigma \partial^{2}c$. These terms arise from contributions involving double contractions with $\sigma\gamma\beta$ in \eqref{psQ} as well as similar contributions related to the anomalous behaviour of $J$, displayed in eqns. \eqref{JJope}, \eqref{TJope}. It is intriguing to see that although the total central charge vanishes, the anomalous behavior of $J$ spoils the nilpotency of the BRST operator. It would be very interesting to study this issue further in the future.

In this work we also found vertex operators in different pictures and they turned out to have the same structure as the ones proposed in \cite{Berkovits:2009by} inspired by the ordinary pure spinor formalism. Here, integrated vertex operators emerged naturally from a simple picture raising operation. Unlike the construction discussed in \cite{Berkovits:2009by}, we dressed up all vertices with a delta function that fixed the dependence of the pure spinor variable in terms of external momentum data. Such a construction relied on the redundant symmetry arising from the twistor-like constraint $(\lambda\gamma^{m})_{\alpha}P_{m} = 0$. The vertices so constructed then resemble those proposed in \cite{Geyer:2019ayz}, and it would thus be interesting to study if the ideas presented in this work could lead to a better understanding of the BRST and critical structure of the ten-dimensional model presented in \cite{Geyer:2019ayz}.

A correlation function measure was then defined in the usual way by integrating over the zero modes of the worldsheet twistor fields and quotienting by the respective symmetry groups of the model. The integration over the modes of $\mu_{a}$ then yielded the 10-dimensional momentum conservation delta function, and the delta functions \eqref{deltafunction} realized the usual scattering equations. Likewise, the polarization and momentum dependence of the correlator \eqref{amplitudeprescription} straightforwardly followed from the close relation between the standard pure spinor formalism and the model constructed here. As a result, the heterotic pure spinor twistor correlator \eqref{amplitudeprescription} gives the same results as the ones obtained from ordinary RNS or pure spinor ambitwistor strings.

The generalization to Type IIB is immediate. The only change to do is to replace the current algebra system by a twistor superfield depending on $\hat{\Gamma}^{m}$ \eqref{psambiactiontypeIIB} and different polarization vectors and spinors. NS-NS, NS-R, R-NS, R-R states are then obtained from the tensor product of the two twistor superfields, and the equivalence of the amplitude \eqref{amplitudeprescription} with standard results easily follows from the fermionic measures for $\Gamma^{m}$ and $\hat{\Gamma}^{m}$ as in \eqref{measureforGamma}. One might also try to describe further CHY models using different matter systems in the pure spinor twistor action \eqref{psambiaction} in a similar manner as done in \cite{Casali:2015vta}. We pretend to explore this further in future work.


\acknowledgments
We are thankful to Nathan Berkovits for discussions and comments on the draft. M.G. would like to thank Renann Jusinskas, Oliver Schlotterer and Lionel Mason for valuable and enlightening discussions. D.G.S would like to thank the Abdus Salam International Centre for Theoretical Physics, ICTP-SAIFR/IFT- UNESP, FAPESP grant 2016/01343-7, CAPES-PROEX, and Perimeter Institute for partial financial support. M.G. was supported by the European Research Council under ERC-STG-804286 UNISCAMP. This research was supported in part by Perimeter Institute for Theoretical Physics. Research at Perimeter Institute is supported by the Government of Canada through the Department of Innovation, Science, and Economic Development, and by the Province of Ontario through the Ministry of Research and Innovation.

\appendix

\section{OPEs for Interacting 2D CFTs}  \label{AppendixA}

In the RNS superstring, or in other standard cases, the worldsheet variables defining the model have OPEs corresponding to those of free fields; that is, OPEs with a single singular term whose coefficient is a constant. In this case one can regularize the product of two fields (at the same point) just by subtracting the corresponding expectation value. However, this procedure fails for fields whose OPEs do not take the form just mentioned. This is case when we try to regularize, for example, $T_{(z)} T_{(w)}$ as $z \rightarrow w$. The above procedure consisting of subtracting $\langle T_{(z)} T_{(w)} \rangle$ removes the most singular term (the central charge term), but the other two singularities remain. In our context, we do not have free OPEs due to the presence of the constraints defining our variables. In virtue of the pure spinor constraint, for instance:
\begin{equation}
    \barmu_{\alpha (z)} \lambda^{\beta}_{(w)} = -\frac{\big(\delta_{\alpha}^{\beta} - K_{\alpha}^{\ \beta}\big)_{(w)}}{(z-w)} + \cdots,
\end{equation}
where the ellipsis stands for the remaining finite terms of the OPE when $z \rightarrow w$. It is clear what we have to do in the more general case: in order to regularize the product of two fields we just need to subtract all singular terms in the OPE. As we are faced with this situation in the current work, in this Appendix we provide a quick review of the theory of OPEs for interacting 2D CFTs based on \cite{DiFrancesco:1997nk, Bais:1987dc}. Proofs of the statements here stated can be found in those references. \\ 

\textbf{Normal Ordering Definition.} We write the OPE between two operators $A_{(z)}$ and $B_{(w)}$ as:
\begin{equation} \label{OPEofAandB}
    A_{(z)} B_{(w)} = \sum_{n=-\infty}^{N} \frac{\{AB\}_{n (w)}}{(z-w)^{n}} = \sum_{n=1}^{N} \frac{\{AB\}_{n (w)}}{(z-w)^{n}} + \{AB\}_{0 (w)} + \cdots,
\end{equation}
and separate those terms which diverge when $z \rightarrow w$ from those that are finite:
\begin{align}
& \llangle A_{(z)} B_{(w)} \rrangle =  \sum_{n=1}^{N} \frac{\{AB\}_{n (w)}}{(z-w)^{n}}, \label{singularterms} \\[0.5ex]
& A_{(z)} B_{(w)} = \llangle A_{(z)} B_{(w)}\rrangle + \big( A_{(z)} B_{(w)}\big), \label{OPEseparating}
\end{align}
where $\big( A_{(z)} B_{(w)}\big)$ stands for the remaining terms in \eqref{OPEofAandB} that are finite as $z \rightarrow w$. The normal ordered product between two operators at coincident points is then defined as:
\begin{equation} \label{normalorderdef}
    \big( AB \big)_{(z)} = \{AB\}_{0 (z)},
\end{equation}
which can also be expressed as a contour integral in the following way:
\begin{equation} \label{normalorderintegral}
    (AB)_{(w)} = \oint_{w} \frac{\mathrm{d}x}{(x-w)} A_{(x)}B_{(w)}.
\end{equation}
It is straightforward to show the equivalence between this expression and \eqref{normalorderdef} by simply replacing \eqref{OPEofAandB} in \eqref{normalorderintegral}. Using the definition \eqref{singularterms} to extract the singular terms in the $A_{(z)}B_{(w)}$ OPE, we see that our definition of normal ordering satisfies:
\footnote{Compare this with the regularization in free theories consisting of subtracting the corresponding expectation value $:\phi\phi:_{(z)} = \lim_{w \rightarrow z} (\phi_{(z)}\phi_{(w)} - \langle \phi_{(z)} \phi_{(w)} \rangle$), which in term of modes is equivalent to the more familiar normal ordering where annihilation operators are placed in the rightmost position. See chapters 5 and 6 in \cite{DiFrancesco:1997nk} for more details.}
\begin{equation}
    \big(AB\big)_{(w)} = \lim_{z \rightarrow w} \bigg[ A_{(z)}B_{(w)} - \llangle A_{(z)}B_{(w)}\rrangle \bigg].
\end{equation}
The explicit form of the regular terms in \eqref{OPEseparating} can be extracted from the Taylor expansion of $A(z)$ around $w$:
\begin{equation}
    \big( A_{(z)} B_{(w)} \big) = \sum_{n \geq 0} \frac{(z-w)^{n}}{n!} (\partial^{n}A \ B)_{(w)}.
\end{equation}
We stress that this definition of normal-ordering is neither commutative nor associative:
\begin{align}
    (AB) & \neq (BA), \\
    \big(A(BC)\big) & \neq \big((AB)C\big),
\end{align}
and thus some rearrangement lemmas (explained below) are necessary. \\

\textbf{A Generalized Wick Theorem.} It is usually the case that one has to compute OPEs involving composite operators of interacting fields. We thus need a version of Wick's theorem specialized to the current definition of normal ordering which extracts the corresponding divergent terms. This is given by:
\begin{align}
    \llangle A_{(z)} (BC)_{(w)} \rrangle & = \oint_{w} \frac{\mathrm{d}x}{(x-w)} \bigg\{ \llangle A_{(z)}B_{(x)} \rrangle  C_{(w)} + B_{(x)} \llangle A_{(z)} C_{(w)} \rrangle \bigg\} \\[1.0ex]
    & = \oint_{w} \frac{\mathrm{d}x}{(x-w)} \Bigg\{\mathlarger{\sum}_{n>0}\frac{\{AB\}_{n (x)} C_{(w)}}{(z-x)^{n}} + \mathlarger{\sum}_{n > 0} \frac{B_{(x)} \{AC\}_{n (w)}}{(z-w)^{n}} \Bigg\}.
\end{align}
If one needs to calculate $(BC)_{(z)}A_{(w)}$ one first computes $A_{(z)}(BC)_{(w)}$, interchange $z \leftrightarrow w$, and expand the operators at $z$ around $w$. \\

\textbf{Rearrangement Lemmas.} As we previously stressed, when working with the previous type of expressions one must consider the fact that the definition of normal ordering is neither commuting nor associative. Useful formulae in this regard are given by
\begin{align}
   (AB)-(BA) & = ([A,B]), \\
    \big(A (BC)\big) - \big( B(AC) \big) & = \big( ([A,B]) C \big), \\[0.5ex]
    \big( (AB) C \big) - \big(A (BC) \big) & = \big(A ([C,B])\big) + \big( ([C,A]) B \big) + \big( [(AB),C]\big),
\end{align}
which provide some commutation properties and the violation of associativity. The normal ordered field commutator can be expressed in terms of the singular coefficients of the OPE as:
\begin{align}
    & \big( [A,B] \big)_{(w)} = \sum_{n \geq 1} \frac{(-1)^{n+1}}{n!} \partial^{n}\{AB\}_{n (w)}.
\end{align} 

\textbf{An Example.} In order to illustrate the previous considerations let us provide a quick illustration of how to obtain the most singular term in the $J_{B(z)} N^{mn}_{CD(w)}$ OPE for arbitrary correction constants $B$, $C$ and $D$ in
\begin{align}
    & J_{B(z)} = (\barmu_{\alpha} \lambda^{\alpha}) - (\barlambda^{\alpha}\mu_{\alpha}) + (\bar{\Gamma}^{m} \Gamma_{m}) + B (Y_{\alpha} \partial \lambda^{\alpha}), \\[1.0ex] & N^{mn}_{CD(z)} = -\frac{1}{2}(\barmu \gamma^{mn} \lambda) + \frac{1}{2}(\barlambda \gamma^{mn} \mu) - 2 (\bar{\Gamma}^{[m} \Gamma^{n]}) + C(\partial Y \gamma^{mn} \lambda) + D(Y \gamma^{mn} \partial \lambda).
\end{align}
For instance:
\begin{align}
    \big \llangle \barmu_{\alpha} \lambda^{\alpha}_{(z)} -\frac{1}{2}&(\barmu \gamma^{mn} \lambda)_{(w)} \big \rrangle = \nonumber \\&- \frac{(\gamma^{mn})^{\gamma}_{\ \delta}}{2} \oint \frac{\mathrm{d}x}{(x-w)}\big[ \llangle \barmu_{\alpha} \lambda^{\alpha}_{(z)} \barmu_{\gamma (x)} \rrangle \lambda^{\delta}_{(w)} + \barmu_{ \gamma(x)} \llangle \barmu_{\alpha} \lambda^{\alpha}_{(z)} \lambda^{\delta}_{(w)} \rrangle  \big]. \label{example}
\end{align}
It is straightforward to see that the second term does not contribute to the most singular term (the simple pole is the only non-zero contribution). Meanwhile, to compute $\llangle \barmu_{\alpha} \lambda^{\alpha}_{(z)} \barmu_{\gamma (x)} \rrangle$, we first compute $\llangle \barmu_{\gamma (z)} \barmu_{\alpha} \lambda^{\alpha}_{(x)} \rrangle$, interchange $z \leftrightarrow x$, and expand $z$ around $x$. The result is:
\begin{equation}
    \llangle \barmu_{\alpha}\lambda^{\alpha}_{(z)} \barmu_{\gamma (x)} \rrangle = \frac{\big(\barmu_{\alpha}(\delta^{\alpha}_{\gamma}-K_{\gamma}^{\ \alpha})\big)_{(x)}}{(z-x)}.
\end{equation}
In order to properly compute the different pole contributions in the first term we need to separate $\big(\barmu_{\alpha}(\delta^{\alpha}_{\gamma}-K_{\gamma}^{\ \alpha})\big)_{(x)} \lambda^{\delta}_{(w)} $ into its singular and normal ordered components as in \eqref{OPEseparating}. Only the singular component will contribute to the most singular term in \eqref{example}:
\begin{equation}
    \llangle \big(\barmu_{\alpha}(\delta^{\alpha}_{\gamma}-K_{\gamma}^{\ \alpha})\big)_{(x)} \lambda^{\delta}_{(w)} \rrangle = - \frac{((\delta_{\alpha}^{\delta} - K_{\alpha}^{\ \delta})(\delta^{\alpha}_{\gamma} - K_{\gamma}^{\ \alpha}))_{(w)}}{(x-w)}.
\end{equation}
Notice that at this step we only have $\lambda$'s and $Y$'s, so ordering becomes unimportant. When evaluating the double pole integral one finds:
\begin{equation}
    \big \llangle \barmu_{\alpha} \lambda^{\alpha}_{(z)} -\frac{1}{2}(\barmu \gamma^{mn} \lambda)_{(w)} \big \rrangle = \frac{3}{2} \frac{(Y \gamma^{mn} \lambda)_{(w)}}{(z-w)^{2}} + \ldots,
\end{equation}
where the ellipsis stands for poles of lower order. Following a similar procedure with the $\barlambda/\mu$, $\bar{\Gamma}/\Gamma$ pairs, and the correction terms, one finds in general that
\begin{equation}
    \llangle J_{B(z)}N^{mn}_{CD (w)} \rrangle = \Big( 2 + \frac{B}{2} + C - D \Big) \frac{(Y \gamma^{mn} \lambda )_{(w)}}{(z-w)^{2}} + \ldots,
\end{equation}
where the ellipsis again stands for poles of lower order.


\providecommand{\href}[2]{#2}\begingroup\raggedright\endgroup



\end{document}